\documentclass[journal]{IEEEtran}
\IEEEoverridecommandlockouts
\usepackage{cite}
\usepackage{amsmath,amssymb,amsfonts,extpfeil}
\usepackage{algorithmic}
\usepackage{graphicx}
\usepackage{textcomp}
\usepackage{xcolor}
\def\BibTeX{{\rm B\kern-.05em{\sc i\kern-.025em b}\kern-.08em
		T\kern-.1667em\lower.7ex\hbox{E}\kern-.125emX}}
\usepackage{algorithmic}
\usepackage{algorithm}
\usepackage{subfigure}
\usepackage{lettrine}

\renewcommand{\IEEEPARstart}[2]{\lettrine[findent=3pt,nindent=0pt,lines=2,lraise=0.05,lhang=0.]{#1}{#2}}\usepackage{caption}
\makeatletter

\newcommand{\Rmnum}[1]{\expandafter\@slowromancap\romannumeral #1@}
\makeatother
\newcommand{\mv}[1]{\mbox{\boldmath{$ #1 $}}}
\usepackage[english]{babel}
\usepackage{amsmath}
\usepackage{graphicx} 
\usepackage{epstopdf}
\usepackage{amssymb}
\usepackage{amsfonts}
\usepackage{amsthm}
\usepackage{cite}
\usepackage{bm,comment}

\usepackage{subeqnarray}
\usepackage{subfigure}
\usepackage{multicol}
\usepackage{multirow}
\usepackage{diagbox}
\usepackage{slashbox}
\usepackage{stfloats}
\usepackage{float}
\usepackage{color} 
\usepackage{cases}

\newtheorem{remark}{Remark}

\graphicspath{{figs/}}

\begin{document}
	
	\setlength{\lineskiplimit}{0pt}
	\setlength{\lineskip}{0pt}
	\setlength{\abovedisplayskip}{3pt}   
	\setlength{\belowdisplayskip}{3pt}
	\setlength{\abovedisplayshortskip}{3pt}
	\setlength{\belowdisplayshortskip}{3pt}

	\title{Movable Antenna Enhanced Multi-Region Beam Coverage: A Multi-Notch-Filter-Inspired Design}
 	\author{Dong Wang, \IEEEmembership{Student Member, IEEE}, Weidong Mei, \IEEEmembership{Member, IEEE},  Zhi Chen, \IEEEmembership{Senior Member, IEEE}, Boyu Ning, \IEEEmembership{Member, IEEE}	
    \vspace{-0em}
	\thanks{
		The authors are with the National Key Laboratory of Wireless Communications, University of Electronic Science and Technology of China, Chengdu 611731, China (e-mail: DongwangUESTC@outlook.com; wmei@uestc.edu.cn; chenzhi@uestc.edu.cn; boydning@outlook.com).
        }}
	\maketitle
	
	\begin{abstract}
        Movable antenna (MA) has emerged as a promising technology to enhance wireless communication performance by exploiting the new degree of freedom (DoF) via antenna position optimization. In this letter, we investigate the MA-enhanced wide beam coverage over multiple subregions in the spatial domain. Specifically, we aim to maximize the minimum beam gain over the desired subregions by jointly optimizing the transmit beamforming and antenna position vector (APV).  Although this problem is non-convex, we propose an efficient algorithm to solve it by leveraging the similarity between the considered multi-region coverage and classical multi-notch filter (MNF) design. In particular, we construct a spatial MNF-based transmit beamforming vector by assuming a continuous amplitude and phase-shift profile within the antenna movement region. Based on this continuous profile, we propose a sequential update algorithm to select an optimal subset of MA positions for multi-region coverage, jointly with a Gibbs sampling (GS) procedure to avoid undesired local optimum. Numerical results show that our proposed algorithm can significantly outperform conventional fixed position antennas (FPAs) and achieve a comparable performance to the alternating optimization (AO) algorithm with dramatically lower complexity.
        
		\begin{IEEEkeywords}
			Movable antenna, wide-beam coverage, antenna position optimization, multi-notch filter, Gibbs sampling.
		\end{IEEEkeywords}     
	\end{abstract}
	\vspace{-1em}
	\begingroup
	\allowdisplaybreaks
	\section{Introduction}

    \IEEEPARstart{T}{he} multi-antenna technology has been a cornerstone in the evolution of wireless communications, which exploits spatial degrees of freedom (DoFs) to achieve substantial gains in both energy and spectral efficiency  \cite{MIMO1}.
     However, conventional multi-antenna technologies rely on fixed-position antennas (FPAs) at the transmitter (Tx) and/or receiver (Rx), which limit their adaptability to dynamic wireless channels and flexibility in beamforming.
    
    To harness the continuous spatial DoFs in wireless channels, the movable antenna (MA) technology has emerged as a promising solution. Compared to conventional FPAs, MAs enable local antenna movement within a confined region at the Tx/Rx, which gives rise to a strong capability for multi-path channel reshaping and flexible beamforming \cite{MA}.     
    In light of these advantages, there have been extensive studies devoted to MA position optimization for multi-channel reshaping in various scenarios, e.g., multiple-input single-output (MISO) \cite{6dma,p2p}, integrated sensing and communication (ISAC) \cite{isac}, cognitive radio \cite{cognitive_radio}, physical-layer security \cite{ref_phy1,ref_phy2}, wideband communications \cite{wideband}, intelligent reflecting surface (IRS)-aided communications \cite{irs}, among others.

    \begin{figure}[t]
		\centering
        \includegraphics[width=5.2cm,height=3.2cm]{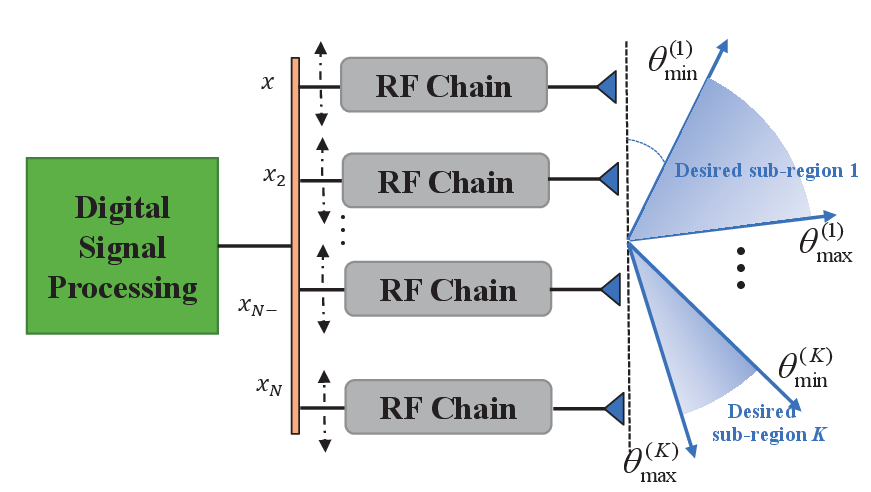}
		\caption{MA-enhanced multi-region beam coverage.}\label{sys_model}
		\vspace{-20pt}
    \end{figure}

	Furthermore, by properly adjusting the MAs' positions, they can also alter the spatial correlation among steering vectors corresponding to different angles, thereby achieving more efficient array signal processing, such as beam nulling \cite{nulling}, multi-beam forming \cite{multi_beam,multi_beam2}, and wide-beam coverage \cite{wang_globecom,twc_fmcw}. In particular, in \cite{wang_globecom} and \cite{twc_fmcw}, the authors aimed to jointly optimize the transmit beamforming and antenna positions to maximize the worst-case beam gain within one or multiple continuous regions in the angular domain. Due to the non-convex nature of this optimization problem, the authors proposed an alternating optimization (AO) algorithm to obtain a suboptimal solution. Although this algorithm can achieve effective wide-beam coverage, the resulting complexity order is polynomial and becomes practically unaffordable for a large number of MAs or antenna movement region. In the case of single-region coverage, the authors in \cite{twc_fmcw} proposed a lower-complexity MA position optimization algorithm based on the frequency modulation continuous wave (FMCW). However, this algorithm cannot be applied to the scenario of multi-region coverage or digital beamforming architecture.
		
    In view of this fact, this letter aims to develop an efficient antenna position optimization algorithm for multi-region wide beam coverage with MAs.  Specifically, we aim to maximize the minimum beam gain over multiple angular regions by jointly optimizing the transmit beamforming and the antenna position vector (APV). Different from the prior works relying on AO, we propose a more efficient algorithm by drawing an analogy between the considered multi-region beam coverage problem and the classical multi-notch filter (MNF) design (also referred to as multi-band-stop filter). In particular, we construct a spatial MNF-based transmit beamforming vector by assuming a continuous amplitude and phase-shift profile within the antenna movement region. Based on this continuous profile, we propose a sequential update algorithm to select an optimal subset of MA positions to approximate the ideal beam coverage performance achieved by an MNF. To avoid undesired local optimum,  a Gibbs sampling (GS) procedure is added between two consecutive sequential update rounds. Numerical results demonstrate that our proposed algorithm substantially outperforms conventional FPAs and achieve a comparable performance to the AO algorithm with dramatically lower complexity.

    {\it Notations:}  $\mathcal{A}\cup\mathcal{B}$ and $\mathcal{A}\backslash\mathcal{B}$ denote the union and subtraction sets of $\mathcal{A}$ and $\mathcal{B}$, respectively.
	For a complex-valued vector $\mv{x}$, $\mv{x}^T$, $\mv{x}^H$, $\Vert\mv{x}\Vert_2$, $\mv{x}(n)$  and $\text{vec}(\mv{x})$ denote its transpose, conjugate transpose, $l_2$-norm, the $n$-th element and vectorized version, respectively. For a matrix $\mv{A}$, $\text{Tr}(\mv{A})$ and $\mv{A}(m,n)$ represent its trace and the $(m,n)$-th entry, respectively.
   \begingroup
	\allowdisplaybreaks     
	
	\vspace{-5pt}
    \section{System Model}

As shown in Fig. 1, we consider a Tx equipped with a linear array with $N$ MAs, whose positions can be flexibly adjusted within a 1D line segment denoted as $\mathcal{C}_t$. The length of the line segment is denoted as $D$. Note that the proposed method in this paper can also be extended to a planar MA array by decomposing the overall beam gain into its horizontal and vertical components.
For convenience, we assume a reference origin point $x_0=0$ in $\mathcal{C}_t$, as shown in Fig. 1. Let $x_n$ denote the coordinate of the $n$-th MA's position, $n\in \mathcal N \triangleq \{1,2,\cdots,N\}$. Accordingly, the APV of all $N$ MAs is denoted by $\mv{x}\triangleq[x_1,x_2,\cdots,x_N]^T\in \mathbb{R}^{N\times1}$. Let $f_c$ and $\lambda$ denote the operating carrier frequency and wavelength, respectively, where $f_c=\frac{c}{\lambda}$ with $c$ denoting the velocity of light. For a given steering angle $\theta$, $0\le\theta\le\pi$, the array response vector of the MA array can be determined as a function of the APV $\mv{x}$ and $\theta$, i.e.,
	\begin{align}	
		\mv{a}(\mv{x},\theta) = [e^{j\frac{2\pi}{\lambda}x_1\cos\theta},\cdots,e^{j\frac{2\pi}{\lambda}x_N\cos\theta}]^T\in \mathbb{C}^{N\times 1}. 
	\end{align}
    The transmit beamforming is expressed as
    \begin{align}
        \mv{\omega}=[\alpha_1 e^{j\phi_1},\alpha_2 e^{j\phi_2},\cdots,\alpha_N e^{j\phi_N}]^T\in\mathbb{C}^{N\times 1},
    \end{align}
    where $\alpha_n$ and $\phi_n$ denote the amplitude and phase shift of the $n$-th MA. We assume a unity transmit power in this letter, i.e., $\Vert\mv{\omega}\Vert_2^2=1$. As such, the beam gain toward the steering angle $\theta$ is given by
    \begin{align}
        G(\mv{\omega},\mv{x},\theta)=\big\vert\mv{\omega}^H\mv{a}(\mv{x},\theta)\big\vert^2.\label{gain_initial}
    \end{align}

    As shown in Fig. 1, we aim to achieve uniform beam coverage in the angular domain by jointly optimizing the APV $\mv{x}$ and the transmit beamforming $\mv{\omega}$. In particular, we consider that the desired coverage region is composed of $K$ disjoint subregions, i.e., $\mathcal{R}=\mathcal{R}_1\cup\mathcal{R}_2\cup\cdots\cup\mathcal{R}_K$, with $\mathcal{R}_k=\{\theta\in[\theta_{\min}^{(k)},\theta_{\max}^{(k)}]\}$, where $\theta_{\min}^{(k)}$ and $\theta_{\max}^{(k)}$ denote the boundary angles of the $k$-th subregion, with $\theta_{\min}^{(k)}\le \theta_{\max}^{(k)}$. By changing the number of the subregions (i.e., $K$) and the width of each subregion $k$ (i.e., $\theta_{\max}^{(k)}-\theta_{\min}^{(k)}$), flexible beam coverage can be achieved. 

    To achieve uniform beam gain over the desired $K$ subregions, we aim to maximize the minimum beam gain over them, i.e.,
    \begin{align}     
    G_{\min}(\mv{\omega},\mv{x})=\min_{\theta\in\mathcal{R}}G(\mv{\omega},\mv{x},\theta).
    \end{align}
    The associated problem can be formulated as    
    \begin{subequations}
        \begin{align}
            \text{(P1)} \quad&\max_{\mv{\omega},\mv{x}}\,G_{\min}(\mv{\omega},\mv{x})\label{p1obj}\\
            \text{s.t.} \quad & 0\leq x_n\leq D, \quad n\in\mathcal{N},\label{apv_cons1}\\
            & \vert x_i-x_j\vert\geq d_{\min}, \quad \forall i\neq j, i,j\in\mathcal{N},\label{apv_cons2}\\
            & \Vert\mv{\omega}\Vert_2^2= 1\label{abf_cons}.
        \end{align}
    \end{subequations}
    where (\ref{apv_cons1}) ensures that the MAs are located within the transmit region $\mathcal{C}_t$, and (\ref{apv_cons2}) ensures a minimum distance between any two MAs (denoted as $d_{\min}$) to avoid antenna coupling. Notably, the proposed multi-region wide-beam coverage scheme can be applied during the initial access stage to establish reliable wireless connections between the BS and users uniformly distributed within the desired regions.\footnote{It should be mentioned that since the user distribution (i.e., $\theta_{\min}^{(k)}$, 
    $\theta_{\max}^{(k)}$, $k=1,2,\cdots,K$) varies slowly over time, the APV in this stage can be updated at a low frequency.}
 Once their instantaneous channel information becomes available, the transmit beamforming and antenna positions can be re-optimized to maximize practical communication metrics such as the signal-to-interference-plus-noise ratio (SINR).
    
    However, (P1) is generally difficult to be optimally solved due to the coupling between the transmit beamforming $\mv{\omega}$ and APV $\mv{x}$ in the objective function. In our previous work \cite{wang_globecom,twc_fmcw}, we have proposed an AO algorithm to solve it suboptimally. Its basic idea is to alternately optimize $\mv{\omega}$ and $\mv{x}$ with the other being fixed. For both transmit beamforming and APV optimization subproblems, they can be efficiently solved by invoking the successive convex approximation (SCA) technique. The details are omitted in this letter due to the space limit. Despite the efficacy of the AO algorithm, its computational complexity is given by $\mathcal{O}(\sqrt{L}N(N^2+L))$, which is polynomial with respect to (w.r.t.) $N$ and hence, becomes practically unaffordable. In the following, we propose a more efficient solution to (P1) inspired by the MNF design.

  	\vspace{-8pt}
    \section{Proposed MNF-Inspired Solution to (P1)}
     \begin{figure}[t]
    	\centering
    	\includegraphics[width=4cm,height=3cm]{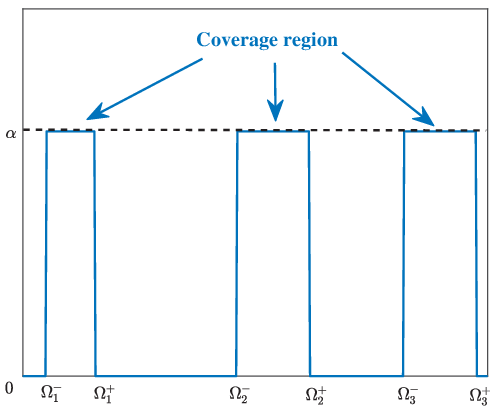}
    	\caption{Frequency/spatial spectrum of an ideal MNF.}\label{band_pass}
    	\vspace{-18pt}
    \end{figure}
        In this section, we present an alternative solution to (P1) by utilizing the principle of MNFs. In particular, we first assume a continuous aperture over the 1D line segment $\mathcal{C}_t$ with a continuous amplitude $\alpha(x)$ and a continuous phase profile $\psi(x)$ at any position $x$, $x\in\mathcal{C}_t$ with $\int_{0}^{D}\alpha^2(x){\rm d} {x=1}$. The array response at any position $x$ toward the steering angle $\theta$ is given by $a(x,\theta)=e^{j\frac{2\pi}{\lambda}x\cos\theta}$. As such, the beam gain toward any steering angle $\theta$ for this continuous array is given by
    \begin{align}
        \mathcal{G}(\theta)=\bigg\vert\int_{0}^{D}\alpha(x)e^{j\psi(x)}a^*(x,\theta){\rm d} x\bigg\vert^2.\label{G_theta}
    \end{align}
    To obtain more insights, define $\Omega=\frac{2\pi}{\lambda}\cos\theta$, $\theta\in\mathcal{R}$. As $\cos\theta$ monotonically decreases with $\theta$ within $[0,\pi]$, it must hold that $\Omega\in\mathcal{I}\triangleq \mathcal{I}_1\cup\cdots\cup\mathcal{I}_K$, where $\mathcal{I}_k=[\Omega_k^-,\Omega_k^+]$, with $\Omega_k^-=\frac{2\pi}{\lambda}\cos\theta_{\max}^{(k)}$ and $\Omega_k^+=\frac{2\pi}{\lambda}\cos\theta_{\min}^{(k)}$. Define an indicator function $\mathbb{I}(x)$  for any position $x$, which is equal to one if $x\in[0,D]$ and zero otherwise. Thus, we can recast $\mathcal{G}(\theta)$ in (\ref{G_theta}) as   
    \begin{align}
        \mathcal{G}(\Omega)=\bigg\vert\int_{-\infty}^{+\infty}\underbrace{\alpha(x)e^{j\psi(x)}\mathbb{I}(x)}_{\triangleq \tilde{\omega}(x)}e^{-j\Omega x}{\rm d} x\bigg\vert^2\triangleq\big\vert \mathcal{\tilde G}(\Omega)\big\vert^2,\label{ft_g_omega}
    \end{align}
    where $\mathcal{\tilde G}(\Omega)=\int_{-\infty}^{+\infty}\tilde\omega(x)e^{-j\Omega x}{\rm d} x$. It is interesting to note that $\mathcal{\tilde G}(\Omega)$ can be interpreted as the Fourier transform (FT) of $\tilde\omega(x)$. The only difference lies in that the integral in (\ref{ft_g_omega}) is taken over the spatial domain instead of the time domain as in FT. Hence, to achieve uniform beam gain over the desired coverage region $\mathcal{R}$, it is desired to find  $\tilde\omega (x)$ that renders $\vert\mathcal{\tilde G}(\Omega)\vert$ as uniform as possible within the interval $\mathcal{I}$.

    To this end, we construct an MNF in the spatial domain as $\mathcal{\tilde G}(\Omega)$ to obtain $\tilde{\omega}(x)$. The MNF has a flat amplitude in each sub-interval $\mathcal{I}_k$, $k=1,\cdots,K$, while the amplitude is zero in other intervals as shown in Fig. \ref{band_pass}. As such, an MNF can be expressed as     
    \begin{align}
        \mathcal{\tilde G}(\Omega)=\mu\sum_{k=1}^{K}\big(u(\Omega-\Omega_k^-)-u(\Omega-\Omega_k^+)\big),\label{band_pass_G}
    \end{align}
    where $\mu$ represents an amplitude coefficient, and $u(\Omega)$ denotes the unit-step function, i.e.,    
     \begin{align}u(\Omega)=\left\{\begin{aligned}
    &1, \quad\text{if} \quad \Omega\geq 0,\\
    &0, \quad\text{otherwise}.
    \end{aligned}\right.\end{align}
    By performing the inverse Fourier transform (IFT) on (\ref{band_pass_G}), we can obtain an ideal $\tilde{\omega}(x)$ given by
    \begin{align}
        &\tilde\omega(x)=\frac{1}{2\pi}\int_{-\infty}^{+\infty}\mu\sum_{k=1}^{K}\big(u(\Omega-\Omega_k^-)-u(\Omega-\Omega_k^+)\big)e^{j\Omega x}\rm{d}\Omega\nonumber\\
       & =\sum_{k=1}^{K}\frac{\mu(\Omega_k^+-\Omega_k^-)}{2\pi}{\rm sinc}\left(\frac{(\Omega_k^+-\Omega_k^-)}{2}x\right)e^{j\frac{\Omega_k^{-}+\Omega_k^+}{2}x},
    \end{align}
    where ${\rm sinc}(x)=\frac{\sin x}{x}$. To satisfy the unit-power constraint in (\ref{abf_cons}), the ideal beamforming amplitude for this continuous aperture can be obtained as
    \begin{align}
        \vert\omega(x)\vert=\frac{\vert\tilde \omega(x)\vert}{\sqrt{\int_{0}^{D}\vert\tilde\omega(x)\vert^2{\rm d}x}}\triangleq\alpha(x),\label{con_ampli}
    \end{align}
    which depends on antenna position $x$. Furthermore, the ideal beamforming phase shift can be expressed as $\psi(x)=\arg\big(\tilde\omega(x)\big)$. Based on the above, the ideal transmit beamforming is given by
    \begin{align}
        \omega(x)=\frac{\vert\tilde \omega(x)\vert}{\sqrt{\int_{0}^{D}\vert\tilde\omega(x)\vert^2{\rm d}x}}e^{j\arg\big(\tilde\omega(x)\big)},\label{con_omega_x}
    \end{align}
    which can achieve a flat beam gain within the considered region $\mathcal{R}$ (or interval $\mathcal{I}$). However, the above amplitude and phase profile require a continuous aperture from $0$ to $D$, which is practically infeasible. Fortunately, by leveraging the flexible position selection enabled by MAs, it is expected that a close performance to the continuous array in (\ref{con_omega_x}) can be achieved. In particular, given the position of the $n$-th MA, i.e., $x_n$, the MNF-based amplitude and phase shift are given by
    \begin{align}
        \alpha(x_n)=\frac{\vert\tilde\omega(x_n)\vert}{\Vert \mv{\alpha}(\mv{x})\Vert_2} \quad {\rm and}\quad
        \psi(x_n)=\arg\big(\tilde\omega(x_n)\big),\label{ma_ampli_phase}
    \end{align}
    respectively, where $\mv{\alpha}(\mv{x})=[\tilde\omega(x_1),\cdots,\tilde\omega(x_N)]^T$. Then, the MNF-based transmit beamforming vector for all MAs can be determined as
    \begin{align}
        \mv{\omega}(\mv{x})\!\!=\!\!\bigg[\!\!\frac{\vert\tilde\omega(x_1)\vert}{\Vert \mv{\alpha}(\mv{x})\Vert_2}\!e^{j\!\arg\!\big(\tilde\omega(\!x_1\!)\!\big)},\!\!\cdots\!,\!\frac{\vert\tilde\omega(x_N)\vert}{\Vert \mv{\alpha}(\mv{x})\Vert_2}\!e^{j\!\arg\!\big(\tilde\omega(\!x_N\!)\big)\!}\!\bigg]^T.\!\!\label{bpf_dpf_v}
    \end{align}

    Next, by substituting (\ref{bpf_dpf_v}) into (\ref{gain_initial}), the beam gain toward the angle $\theta$ can be expressed as a function of APV $\mv{x}$ as
    \begin{align}
        G(\mv{\omega}(\mv{x}),\mv{x},\theta)&=\big\vert\mv{\omega}(\mv{x})^H\mv{a}(\mv{x},\theta)\big\vert^2\label{gain_bpf}
       \\
        &=\bigg\vert\sum_{n=1}^N \frac{\tilde\omega(x_n)}{\Vert\mv{\alpha}(\mv{x})\Vert_2}e^{j\big(\frac{2\pi}{\lambda}\cos\theta-\arg\big(\tilde\omega(x_n)\big)\big)}\bigg\vert^2.\nonumber
    \end{align}

    Based on (\ref{gain_bpf}), we can formulate the following MNF-based MA position optimization problem, i.e.,
    \begin{subequations}
        \begin{align}
            \text{(P2)} \quad&\max_{\mv{x}}\,G_{\min}(\mv{\omega}(\mv{x}),\mv{x})\label{p1obj}\\
            \text{s.t.} &\quad (\text{\ref{apv_cons1}}),(\text{\ref{apv_cons2}}).
        \end{align}
    \end{subequations}
    Note that compared to (P1), (P2) is more tractable without the coupling between the beamforming vector $\mv{\omega}$ and the APV $\mv{x}$. However, it is still difficult to derive the optimal solution to (P2) due to the complex beam-gain expression in (\ref{gain_bpf}) w.r.t. $\mv{x}$. Next, we propose a sequential update algorithm to obtain a high-quality sub-optimal solution to (P2).
	\begin{remark}
		Note that our proposed algorithm can also be applied to single-region beam coverage by setting $K=1$ in (\ref{band_pass_G}).  As a result, the MNF in (8) reduces to a band-pass filter (BPF). The corresponding BPF-based phase shift and amplitude can be calculated according to (\ref{ma_ampli_phase}).
	\end{remark}
	\begin{remark}
		The proposed method is also effective if the path loss between the BS and each subregion is accounted for. Let $\beta_k$ denote the path gain between the BS and the $k$-th subregion. Then, the MNF in (\ref{band_pass_G}) can be modified as $\mathcal{\tilde G}(\Omega)=\sum_{k=1}^{K}\mu_k(u(\Omega-\Omega_k^-)-u(\Omega-\Omega_k^+))$, with $\mu_k=\frac{\mu}{\beta_k}$.
	\end{remark}
	
    \vspace{-8pt}
    \section{Proposed Solution to (P2)}
     In this section, we present our proposed solution to (P2) via sequential update and GS. First, to tackle the continuous nature of each subregion $\mathcal{R}_k$, we discretize it into $L_k$ discrete angles, i.e.,
    \begin{align}
        \theta_l^{(k)}=\theta_{\min}^{(k)}+\frac{l^{(k)}-1}{L^{(k)}-1}Z_k,\, l^{(k)}=1,\cdots,L^{(k)},\label{sample}
    \end{align}
    where $L^{(k)}$ denotes the number of the sampling points in $\mathcal{R}_k$, and $Z_k=\theta_{\max}^{(k)}-\theta_{\min}^{(k)}$ denotes the width of $\mathcal{R}_k$. 
    
    Based on the above discretization, we can recast (P2) as a discrete version, i.e.,
    \begin{subequations}
        \begin{align}
            (\text{P2.1})\quad \max_{\mv{x}} \,&\min\limits_{l\in\mathcal{L}_k, k\in\mathcal{K}} G(\mv{\omega}(\mv{x}),\mv{x},\theta_{l}^{(k)})\\
            \text{s.t.}
             \,\,\,& (\text{\ref{apv_cons1}}),(\text{\ref{apv_cons2}}),
        \end{align}
    \end{subequations}
    where $\mathcal{L}_k=\{1,2,\cdots,L^{(k)}\}$ and $\mathcal{K}=\{1,2,\cdots,K\}$. To solve (P2.1), the sequential update algorithm is applied in this letter, which has also been utilized in some existing works for antenna position optimization with an intractable objective function \cite{cognitive_radio,irs}. 
	\vspace{-8pt}
    \subsection{Sequential Update}    
    To facilitate the sequential update, we uniformly discretize the movable region $\mathcal{C}_t$ into $M$ discrete sampling points (with $M \gg N$), with the position of the $m$-th sampling point denoted as $p_m$. Let the set of the sampling points be denoted as $\mathcal{S}_M=\{p_1,p_2,\cdots,p_M\}$ with $p_{m}< p_{m+1}$, $m=1,2,\cdots,M$. Let $\mv{u}=[u_1,u_2,\cdots,u_N]^T$ denote the index vector of the MAs, i.e., $\mv{x}=[p_{u_1},p_{u_2},\cdots,p_{u_N}]^T$. It follows that finding an optimal $\mv{x}$ is equivalent to finding an optimal index vector $\mv{u}$. Consequently, a discrete optimization problem (P2.2) is obtained by replacing $\mv{x}$ in (P2.1) with $\mv{u}$, i.e.,
    \begin{subequations}
    \begin{align}
        (&\text{P2.2}) \quad \max_{\mv{u}}\, F(\mv{u})\label{obj_value}\\
        \text{s.t.} \,\,&\vert p_{u_i}-p_{u_j}\vert\geq d_{\min},\, \forall i\neq j, i,j\in\mathcal{N}\label{apv_cons3},
    \end{align}    
    \end{subequations}
   where $F(\mv{u})=\min_{l,k}\{G\big(\mv{\omega}(\mv{u}),\mv{u},\theta_l^{(k)}\big)\}$. To address (P2.2), we perform the sequential update in multiple rounds, each including $N$ iterations that sequentially update the positions of the $N$ MAs. Let $\mv{u}^{(i-1)}=[u_{1}^{(i-1)},\cdots,u_{N}^{(i-1)}]^T$ denote the antenna index vector output by the $(i-1)$-th round of the sequential update. Then, in the $i$-th round, the updated position for the $n$-th MA is given by
    \begin{align}       
    	u_{n}^{(i)}=\arg\max_{u_n\in\psi_n^{(i)}}\,F\big(\tilde{\mv{u}}^{(i)}(u_n)\big),\,n\in\mathcal{N},\label{obj_su}
    \end{align}
    where $\tilde{\mv{u}}^{(i)}(u_n)=$
    \begin{equation}
        \begin{dcases}
         [u_1,u_{2}^{(i-1)},u_{3}^{(i-1)},\cdots,u_{N}^{(i-1)}]^T, &{\text{if}}\;n=1,
        \\
        [u_{1}^{(i)},u_{2}^{(i)},\cdots,u_{N-1}^{(i)},u_N]^T,&{\text{if}}\;n=N,
        \\
        [u_{1}^{(i)},\cdots,u_{n-1}^{(i)},u_n,u_{n+1}^{(i-1)},\cdots,u_{N}^{(i-1)}]^T,&{\text{otherwise}}\;,\nonumber
        \end{dcases}
    \end{equation}
    and $\psi_{n}^{(i)}$ denotes the feasibility set of $u_n$ given $\mv{u}=\tilde{\mv{u}}^{(i)}(u_n)$ in (P2.2). After $u_N^{(i)}$ is obtained, we can obtain ${\mv u}^{(i)}=[u_1^{(i)},\cdots,u_N^{(i)}]^T$, and the $(i+1)$-th round of sequential update follows, if convergence is not reached. However, the sequential update algorithm may converge to an undesired local optimal solution. To further improve its performance, we propose additional GS procedures between two consecutive rounds of sequential updates to explore more feasible solutions, thereby circumventing local optimality. 
\vspace{-10pt}
\subsection{Gibbs Sampling} \label{gs} 
    The key steps of the GS is to generate some adjacent and random positions based on the current MA positions for solution exploration. Specifically, after the $i$-th round of the sequential update, we take $\mv{u}^{(i)}$ as the input of the GS, denoted as $\mv{u}_{GS}^{(0)}=\mv{u}^{(i)}$. The GS is performed in $T$ rounds, and each round comprises $N$ iterations corresponding to the $N$ MAs. Let $\mv{u}_{GS}^{(t)}=[u_{GS,1}^{(t)},u_{GS,2}^{(t)},\cdots,u_{GS,N}^{(t)}]^T$ denote the output of the $t$-th GS round and $\mathcal{C}_{GS}(t)=\{\mv{u}_{GS}^{(0)},\mv{u}_{GS}^{(1)},\cdots,\mv{u}_{GS}^{(t-1)}\}$ denote the solution set after this round, $1\leq t\leq T$. 
    
   In the $n$-th iteration of the $t$-th GS round, we generate $S$ candidate positions for exploration, including $S_1$ adjacent positions to $u_{GS,n}^{(t-1)}$ and $S-S_{1}$ random positions. Among these candidate solutions, we choose one of them as $u_{GS,n}^{(t)}$ based on certain criteria (to be specified later). Each adjacent position is obtained by adding and subtracting the index of the current $n$-th MA, i.e., $u_{GS,n}^{(t-1)}$, by a certain amount (with those of other MAs fixed). Let $J$ denote the maximum index shift. As a result, in the $n$-th iteration, we can obtain the following $2J$ adjacent positions to $u_{GS,n}^{(t-1)}$, i.e.,\\
   	 $\tilde{\mv{u}}_{GS,n}^{(t)}=$
 	\begin{equation}
 		\begin{dcases}
    		[\tilde u_{GS,1}^{(t)}+j,u_{GS,2}^{(t-1)},\cdots,u_{GS,N}^{(t-1)}]^T, &\text{if}\,\, n = 1, \\
     		[u_{GS,1}^{(t)},u_{GS,2}^{(t)},\cdots,u_{GS,N-1}^{(t)},\tilde{u}_{GS,N}^{(t)}+j]^T, &\text{if}\,\, n = N,
			\\
	  		[u_{GS,1}^{(t)},\cdots,u_{GS,n-1}^{(t)},\tilde{u}_{GS,n}^{(t)}+j,\cdots,u_{GS,N}^{(t-1)}]^T, &\text{otherwise},
		\end{dcases} 
		\nonumber
	\end{equation} 
	with $j\in\mathcal{J}=\{-J,-(J-1),\cdots,J-1,J\}$. Note that some of the above adjacent solutions may not satisfy the minimum distance constraints in (\ref{apv_cons3}). Hence, we only select $S_1$, $S_1\leq 2J$ feasible adjacent solutions from them. Moreover, we randomly generate $S-S_1$ random feasible positions for the $n$-th MA. Let $\tilde{\mathcal{U}}_{GS,n}^{(t)}=\{\tilde{u}_{GS,n}^{(t,1)},\cdots,\tilde{u}_{GS,n}^{(t,S)}\}$ denote the candidate solution set including the above adjacent and random positions for $u_{GS,n}^{(t)}$, assuming that they are sorted in an ascending order, i.e., $\tilde{u}_{GS,n}^{(t,s_1)}<\tilde{u}_{GS,n}^{(t,s_2)}$, $1\leq s_1<s_2\leq S$. Moreover, let $\tilde{\mv{u}}_{GS,n}^{(t,s)}$ denote the resulting antenna index vector $\tilde{\mv{u}}_{GS,n}^{(t)}$ by setting its $n$-th entry as ${\tilde u}_{GS,n}^{(t,s)}$.   
    To proceed, we calculate the probability that the $s$-th element in $\mathcal{\tilde U}_{GS,n}^{(t)}$ is selected as $u_{GS,n}^{(t)}$, which is given by \cite{ref_gs}
     \begin{align}
        \mathrm{Pr}(u_{GS,n}^{(t)}=\tilde{u}_{GS,n}^{(t,s)}\vert u_{GS,n}^{(t-1)})=
        \frac{e^{\gamma F(\tilde{\mv{u}}_{GS,n}^{(t,s)})}}{\sum_{s=1}^{S}e^{\gamma F(\tilde{\mv{u}}_{GS,n}^{(t,s)})}},\label{prob}
    \end{align}    
	  where $\gamma>0$ is a fixed parameter. Given the probabilities in (\ref{prob}), we can generate a uniformly distributed random number $p$ between 0 and 1 and determine $u_{GS,n}^{(t)}$ as $u_{GS,n}^{(t)}=\tilde{u}_{GS,n}^{(t,s')}$, where $\sum_{s=1}^{s'-1}\mathrm{Pr}(u_{GS,n}^{(t)}=\tilde{u}_{GS,n}^{(t,s')}\vert u_{GS,n}^{(t-1)})<p<\sum_{s=1}^{s'}\mathrm{Pr}(u_{GS,n}^{(t)}=\tilde{u}_{GS,n}^{(t,s')}\vert u_{GS,n}^{(t-1)})$. As such, by conducting $T$ GS iterations, the GS solution set is obtained as $\mathcal{C}_{GS}(T)=\{\mv{u}_{GS}^{(0)},\mv{u}_{GS}^{(1)},\cdots,\mv{u}_{GS}^{(T)}\}$. Finally, we update the output of the $i$-th sequential update as,
    \begin{align}
        \mv{u}^{(i)} = \arg \max_{\mv{u}\in\mathcal{C}_{GS}(T)}\, F(\mv{u})\label{gs_select}.
    \end{align}
    The proposed sequential update algorithm with GS is summarized in Algorithm 1.
    
	\vspace{-2pt}
    \begin{algorithm}[!h]
		\caption{Proposed Algorithm for Solving (P2.2)}
		\label{dsso}
		\renewcommand{\algorithmicrequire}{\textbf{Input:}}
		\begin{algorithmic}[1]
			\REQUIRE $\mv{u}^{(0)}$, $T$, $N$ and $i=1$. 
		  \WHILE{the convergence is not reached}
            \FOR{$n=1:N$}
            \STATE Calculate $u_n^{(i)}$ according to (\ref{obj_su}).
            \ENDFOR
            \STATE Update $\mv{u}_{GS}^{(0)}\leftarrow \mv{u}^{(i)}$, $\mathcal{C}_{GS}(0)=\{\mv{u}_{GS}^{(0)}\}$.
	       \FOR{$t=1:T$}
                \FOR{$n=1:N$}
                \STATE Generate adjacent and random position set $\tilde{\mathcal{U}}_{GS,n}^{(t)}$.
                \STATE Obtain $u_{GS,n}^{(t)}$ based on (\ref{prob}).
                \ENDFOR
                \STATE Update $\mathcal{C}_{GS}(t)=\mathcal{C}_{GS}(t-1)\cup\{\mv{u}_{GS}^{(t)}\}$.
            \ENDFOR
            \STATE Update $\mv{u}^{(i)}$ based on (\ref{gs_select}).
            \STATE $i\leftarrow i+1.$
			\ENDWHILE
			\STATE Set $\mv{x}^*=[p_{u_1^{(i)}},\cdots,p_{u_N^{(i)}}]^T$ and calculate $\mv{\omega}(\mv{x}^*)$ based on (\ref{bpf_dpf_v}).     
			\RETURN $\mv{x}^*$ and $\mv{\omega}(\mv{x}^*)$.
		\end{algorithmic}
	\end{algorithm}  
	
    Note that if lines 5-15 of Algorithm 1 are removed, it becomes the sequential update algorithm only. The complexity of Algorithm 1 is given by 
    $\mathcal{O}(LMN+LTN)$, which is linear in $L$, $M$ and $N$. Notably, Algorithm 1 eliminates the need for SCA in beamforming optimization as required in AO \cite{wang_globecom}, thus incurring significantly lower computational time. Moreover, since the sequential update algorithm ensures a non-decreasing objective value for (P2.2), and the GS procedure after each sequential update round does not reduce this value, the convergence of Algorithm 1 is guaranteed.

   \vspace{-6pt}
    \section{Numerical Results}
    In this section, we provide numerical results to evaluate the performance of our proposed MA-enhanced multi-region beam coverage scheme. Unless otherwise stated, the simulation parameters are set as follows. The carrier frequency is set to $f_c=1$ GHz. The minimum inter-MA distance is set to $d_{\min}=\lambda/2$. The iteration number of GS is set to $T=50$. The maximum index shift and fixed parameter $\gamma$ in each GS iteration are set to $J=2$ and $\gamma=5$, respectively. The number of candidate solutions in each GS iteration is set to $S=10$. The length of the linear MA array is set to $D=10\lambda$. The number of discrete sample points in sequential update is $M=500$. The number of subregions is set to $K=2$, with $\mathcal{R}_1=[0,20^{\circ}]$ and $\mathcal{R}_2=[150^{\circ},180^{\circ}]$. For performance comparison, we consider the following baseline schemes:

    1) \textbf{FPA}: $N$ FPAs are deployed with half-wavelength spacing, and the transmit beamforming is optimized via the SCA algorithm.

    2) \textbf{Sequential update (SU) only}: In this scheme, the APV are optimized via the sequential update algorithm without GS.
    
    3) \textbf{AO \cite{wang_globecom}}: In this scheme, the transmit beamforming and APV are alternately optimized via AO as in \cite{wang_globecom}.

	4) \textbf{Heuristic algorithms}: The particle swarm optimization (PSO) \cite{pso} and firefly algorithm (FA) \cite{FA} are applied to optimize the APV in (P2), with the hyperparameters set similarly as \cite{pso} and \cite{FA}.

    \begin{figure*}[t]
        \centering
        \subfigure[Optimized beam gain with $N=8$ and $K=2$.]{\includegraphics[width=4.4cm,height=3.3cm]{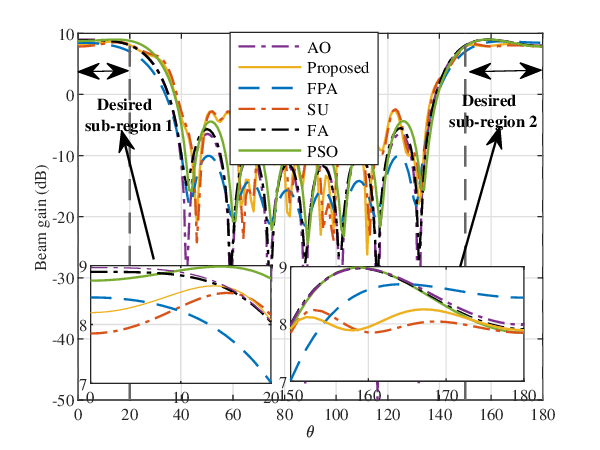}\label{fig_k2}}
        \qquad\quad
        \subfigure[Max-min beam gain versus $N$.]{\includegraphics[width=4.4cm,height=3.3cm]{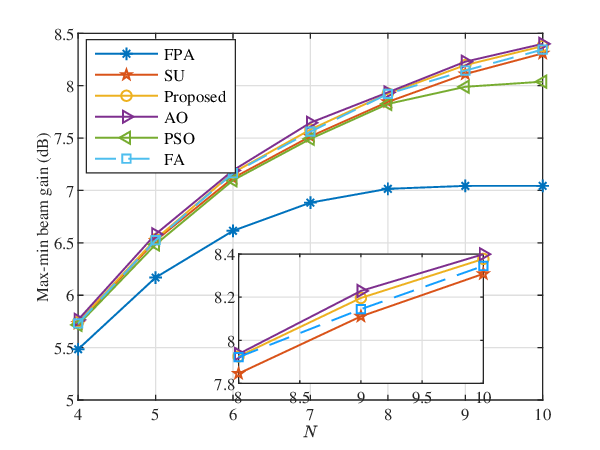}\label{fig_max_min_gain}}
        \qquad\quad
        \subfigure[Execution time by different schemes versus $N$.]{\includegraphics[width=4.4cm,height=3.3cm]{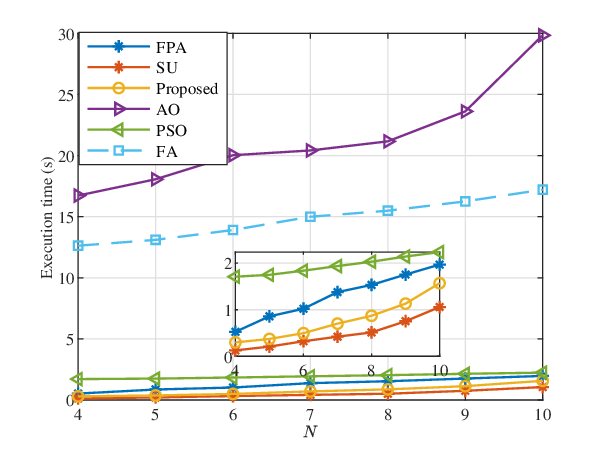}\label{fig_run_time}}
        \vspace{-6pt}
        \caption{Performance evaluation of the proposed algorithm and other baselines.}
        \vspace{-12pt}
    \end{figure*}
		    	
    First, Fig. \ref{fig_k2} plots the optimized beam gains over the coverage region $\mathcal{R}$ by different schemes in the case of $N=8$. It is observed from Fig. \ref{fig_k2} that all schemes with MAs can achieve a more uniform beam coverage and a larger max-min beam gain over $\mathcal{R}$ compared to the FPA scheme. For example, compared to FPAs, the proposed MNF-based algorithm can further increase the max-min beam gain by 1 dB. It is also observed that the max-min beam gains by the proposed algorithms (with and without GS) are close to that by AO. 
        
    Furthermore, we plot the max-min beam gain versus the number of MAs, $N$, in Fig. \ref{fig_max_min_gain}. It is observed that the max-min beam gains by all schemes increase monotonically with $N$. However, the performance of the FPA is observed to saturate as $N$ increases to 8, which implies that less beam gain is exploited for wide-beam coverage for FPAs. This also results in an increasing performance gap between FPAs and MAs as $N$ increases, e.g., 1.5 dB for $N=10$.  It is also observed that the performance gap between the proposed algorithm and the AO algorithm is marginal over the whole range of $N$ considered. Furthermore, employing GS can slightly enhance the max-min beam gain compared to sequential update only. In addition, the proposed algorithm performs slightly better than the PSO and FA under our considered setup. This is mainly due to the less stable performance of PSO and FA, which involve a large number of hyperparameters. Nonetheless, their performance can be improved through more extensive hyperparameter tuning, albeit at the cost of increased complexity.
    
	Finally, Fig. \ref{fig_run_time} plots the execution time by all considered schemes versus $N$. It is observed that although AO achieves the best max-min beam gain among all considered schemes, its runtime is nearly 30 times longer than the proposed scheme, with only a marginal performance gain reaped, as observed from Fig. \ref{fig_max_min_gain}. It is also observed that the GS procedure slightly increases the execution time, e.g., 0.5 s for $N=10$. Interestingly, by dispensing with the need for beamforming optimization, the proposed algorithm even results in a lower execution time than FPA. The reason is that the proposed algorithm only incurs a linear complexity order, while the transmit beamforming design for FPAs incurs a polynomial complexity order. The above observations imply that the proposed algorithm can achieve a more efficient performance-complexity tradeoff than AO. Meanwhile, it is also observed from Fig. \ref{fig_run_time} that both the PSO and FA incur longer execution time than the proposed algorithm.

    \vspace{-2pt}
    \section{Conclusion}
    In this letter, we investigated an MA-enhanced multi-region wide-beam coverage problem, aiming to maximize the minimum beam gain over multiple angular subregions.
    To this end, we proposed an efficient MNF-inspired antenna position optimization algorithm, where the sequential update algorithm was employed jointly with GS to select an optimal subset of MA positions. Numerical results demonstrated that our proposed algorithm can achieve a comparable performance to the AO algorithm with dramatically lower complexity, and significantly outperform  conventional FPAs. It would be interesting to investigate multi-region beam coverage under hybrid beamforming architectures or near-field channel models in future work.

	\vspace{12pt}

\end{document}